\documentclass[aps,pre,reprint,floatfix]{revtex4-1}
\pdfoutput=1

\usepackage[pdftex]{graphicx}
\usepackage{amsmath}
\usepackage{amsfonts}
\usepackage{mathrsfs}
\usepackage{bm}
\usepackage{hyperref}
\usepackage{longtable}
\usepackage{dcolumn}
        \newcolumntype{d}[1]{D{.}{.}{#1}}  
\usepackage{times}

\begin{document}

\title{First-order transitions and thermodynamic properties in the 2D Blume-Capel model: the transfer-matrix method revisited}

\author{Moonjung Jung and Dong-Hee Kim}
\email{dongheekim@gist.ac.kr}
\affiliation{Department of Physics and Photon Science, School of Physics and Chemistry, Gwangju Institute of Science and Technology, Gwangju 61005, Korea}

\begin{abstract}
We investigate the first-order transition in the spin-$1$ two-dimensional Blume-Capel model in square lattices by revisiting the transfer-matrix method. With large strip widths increased up to the size of $18$ sites, we construct the detailed phase coexistence curve which shows excellent quantitative agreement with the recent advanced Monte Carlo results. In the deep first-order area, we observe the exponential system-size scaling of the spectral gap of the transfer matrix from which linearly increasing interfacial tension is deduced with decreasing temperature. We find that the first-order signature at low temperatures is strongly pronounced with much suppressed finite-size influence in the examined thermodynamic properties of entropy, non-zero spin population, and specific heat. It turns out that the jump at the transition becomes increasingly sharp as it goes deep into the first-order area, which is in contrast to the Wang-Landau results where finite-size smoothing gets more severe at lower temperatures.
\end{abstract}

\maketitle

\section{Introduction}
\label{intro}
The Blume-Capel (BC) model~\cite{Blume1966,Capel1966,Capel1967a,Capel1967b} is one of the fundamental models of the tricritical phenomena that have been observed in a variety of systems~\cite{book1984} ranging from multi-component fluids and ferrimagnets to $^3$He-$^4$He mixtures~\cite{Bafi2015}, and ultracold quantum gases~\cite{Shin2008}. The spin-$1$ BC model is described by the Ising Hamiltonian with a crystal field causing spin anisotropy which is written as
\begin{equation}
\mathcal{H} = -J \sum_{\langle i, j\rangle}s_i s_j  + \Delta \sum_i s_i^2 - h \sum_i s_i,
\end{equation}
where spin $s_i$ at site $i$ can take a value of $1$, $-1$, or $0$, and $J$ and $\Delta$ denote the strengths of the ferromagnetic coupling and crystal field, respectively. The summation $\sum_{\langle i, j\rangle}$ runs over all pairs of nearest-neighbor spins. At zero magnetic field $h=0$, the ferromagnetic BC model typically exhibits a tricritical point connecting the first-order transitions occurring in the lower (higher) temperature (crystal field) area and the second-order transitions occurring on the other side of the phase diagram.

The transfer matrix method was one of the earliest numerical tools used to explore the phase diagram of the ferromagnetic spin-$1$ BC model in two dimensions (2D). Beale's seminar work on the 2D BC model~\cite{Beale1986} provided the first transfer matrix calculations in square lattices for the estimation of a few second- and first-order transition points, the tricritical point, and the tricritical eigenvalue exponents. Later, Xavier \textit{et al.}~\cite{Xavier1998} considered the helical boundary conditions to calculate the conformal anomaly at the tricritical point with the estimation of the phase boundary. The nature of the phase transitions in the 2D BC model has been extensively examined also with various other numerical methods which include the Monte Carlo renormalization group~\cite{Landau1981,Landau1986}, histogram method~\cite{Wilding1996,Plascak2013}, Wang-Landau sampling~\cite{Silva2006,Hurt2007,Malakis2009,Malakis2010,Fytas2011,Kwak2015}, simulated tempering~\cite{Valentim2015}, tensor renormalization group~\cite{Yang2016}, replica-exchange Monte Carlo~\cite{Kimura2016}, and quite recently parallel multicanonical simulations~\cite{Zierenberg2017}. 
While we focus on the pure BC model in this work, the effects of randomness have been also extensively studied~\cite{Malakis2009,Malakis2010,Malakis2012,Theodorakis2012,Fytas2013,Fernandes2010,Fernandes2015,Erichsen2017}, where, for instance, the phase diagram was found to be shifted in presence of bond randomness at low temperatures~\cite{Malakis2010}.    
These large numerical efforts now establish a much more solid picture of the phase diagram and the verification of the conjectured exact Ising tricritical exponents~\cite{denNijs1979,Nienhuis1979,Pearson1980,Nienhuis1982} of the 2D BC model, providing an excellent testbed for new numerical methods and strategies such as the machine learning~\cite{Hu2017} to be applied to a tricritical system. 

Probably the most recent addition to the phase diagram of the pure 2D BC model in square lattices is the first-order transition curve in the low temperature area which was obtained in high resolution by the two-parameter Wang-Landau (WL) sampling~\cite{Kwak2015}, simulated tempering~\cite{Valentim2015}, and multicanonical (MUCA) calculations~\cite{Zierenberg2017}. Prior to these advanced Monte Carlo (MC) methods being employed, the available dataset of transition points at low temperatures was mainly from the early transfer matrix (TM) calculations~\cite{Beale1986}. It turned out that the transition points predicted by the TM at low temperatures clearly deviate from the recent WL and MUCA estimates~\cite{Kwak2015,Zierenberg2017}. However, this discrepancy may not imply a true numerical incapability of the TM method in describing the first-order transitions, considered quite limited computational capacity and algorithms available at the time of the previous TM work.

The TM method has a clear advantage over the usual MC methods in the first-order transitions. It is an exact and deterministic method providing straightforward numerics to compute the free energy of a given strip system. It is free from the impenetrable energy barrier issue avoiding which is an important goal of the advanced MC methods such as MUCA~\cite{Berg1991,Berg1992,Janke1992,Janke1998,Zierenberg2015}, WL~\cite{WL1,WL2}, and tempering~\cite{Valentim2015,Fiore2008,Fiore2011} methods. However, because finding the largest eigenvalues required in the TM method is poorly scalable with the strip size, it heavily relies on the efficiency of the matrix algorithm and computational capacity. Thus, given that much increased computing power and more efficient Krylov subspace algorithms are now available for the eigenproblems, it is a proper time to update the TM calculations to reconsider its practical applicability in the first-order area by direct comparison with the recent MC results.  

In this paper, we construct the detailed phase coexistence curve and investigate the thermodynamic properties in the first-order area of the 2D BC model in square lattices by revisiting the transfer matrix method. An infinite-long strip system is considered with finite width increased up to the size of $18$ lattice sites. The transition points are determined in two ways. We first re-examine the finite-size-scaling analysis to correct the data of Ref.~\cite{Beale1986}, and then we search for the spectral crossing of the transfer matrix~\cite{Xavier1998} to find the discontinuous change of the metastable phases in the low temperature area. The resulting first-order transition points show excellent agreement with the estimates of the recent WL simulations by Kwak et al.~\cite{Kwak2015} and MUCA simulations by Zierenberg et al.~\cite{Zierenberg2017}, and our transfer matrix calculations also provide an extended access to even lower temperature area.

We find that the interfacial tension at the phase coexistence linearly increases as it goes deeper into the first-order area. This leads to the rapid development of a kink in the free energy as the strip width increases, providing a pronounced signature of the first-order transition in the thermodynamic properties. We find that finite-size influence is very much suppressed in the observed jumps in the entropy, nonzero spin density, and specific heat, which is compared with the finite-size smoothing found in the previous Wang-Landau results at low temperatures.  

The rest of the paper is organized as follows. In Section~\ref{sec:method}, the numerical procedures of the transfer matrix method are briefly described. In Section~\ref{sec:result1}, our determination of the phase coexistence curve is presented. In Section~\ref{sec:result2}, the first-order signature in the thermodynamical quantities is presented and their finite-size effects are compared with the Wang-Landau calculations. Conclusions are given in Section~\ref{sec:conclusions}.

\section{Transfer matrix method}
\label{sec:method}

We consider 2D regular lattices of a $L \times M$ strip geometry composed of $M$ coupled chains of $L$ sites. Under the periodic boundary conditions, the Hamiltonian can be rewritten in a symmetric form as 
\begin{equation}
\mathcal{H} = \sum_{j=1}^{M} \left[\frac{1}{2}\mathcal{V}_j + \mathcal{W}_{j,j+1} + \frac{1}{2}\mathcal{V}_{j+1}\right] ,
\end{equation}
where the intra- and inter-chain parts $\mathcal{V}$ and $\mathcal{W}$ are
\begin{eqnarray}
\mathcal{V}_j  &=& \sum_{i=1}^L -J s_{i,j} s_{i+1,j} + \Delta s_{i,j}^2 - h s_{i,j} \, ,\\
\mathcal{W}_{j,j+1} &=&  \sum_{i=1}^L -J s_{i,j} s_{i,j+1} \, .
\end{eqnarray}
The partition function is written accordingly as 
\begin{equation}
Z = \sum_{\{s\}} \prod_{j=1}^M e^{-\frac{\beta}{2} \mathcal{V}_j} e^{-\beta \mathcal{W}_{j,j+1}} e^{-\frac{\beta}{2} \mathcal{V}_{j+1}} \equiv \mathrm{Tr} [ \mathbf{T}^M ] ,
\end{equation}
where $\beta\equiv 1/T$ is inverse temperature. The transfer matrix $\mathbf{T}$ can be further decomposed into a product of sparse matrices (for instance, see Refs.~\cite{Domb1960,Thijssen2007}) as
\begin{equation}
\mathbf{T} = \prod_{i=1}^L \exp [-\frac{\beta}{2} a_{i,i+1}] \exp [-\beta b_i ] \exp [-\frac{\beta}{2} c_{i,i+1}].
\end{equation}
where $a$, $b$, and $c$ are given as
\begin{eqnarray}
a_{i,i+1} &=& -J \sigma_i^\prime \sigma_{i+1}^\prime + (\Delta \sigma_i^\prime - h ) \sigma_i^\prime , \\
b_i &=& -J \sigma_i^\prime \sigma_i , \\
c_{i,i+1} &=& -J \sigma_i \sigma_{i+1} + (\Delta \sigma_i - h ) \sigma_i 
\end{eqnarray}
for the component $\langle \sigma_1^\prime \cdots \sigma_L^\prime | \mathbf{T} | \sigma_1 \cdots \sigma_L \rangle$. The periodic boundary conditions are imposed so that $\sigma_{L+1}=\sigma_1$ and $\sigma_{L+1}^\prime = \sigma_1^\prime$.

In the limit of infinite $M$, the largest eigenvalues of the transfer matrix $\mathbf{T}$ essentially determine the properties of systems. The free energy density $f_L$ can be evaluated in terms of the largest eigenvalue $\lambda_1$ as
\begin{equation}
f_L (T,\Delta,h)= - \frac{T}{L} \ln \lambda_1(T,\Delta,h).
\end{equation}
We calculate thermodynamic quantities through its numerical derivatives. The entropy density $s_L \equiv - \frac{\partial f_L}{\partial T} \big\vert_{\Delta,h}$, non-zero spin population density $q_L \equiv \langle s^2 \rangle = \left(\frac{\partial f_L}{\partial \Delta}\right)_{T,h}$, specific heat $c_L \equiv - T \left( \frac{\partial^2 f}{\partial T^2} \right)_{\Delta,h}$ are computed by using the higher-order finite difference method with the step size of $10^{-8}$ in the five-point stencil. In all calculations, the ferromagnetic coupling $J$ is set to be unity.  

In the transfer matrix analysis, the phase transitions are indicated by the asymptotic degeneracy between the largest eigenvalues of the transfer matrix. The correlation length $\xi_L \equiv 1/\ln(\lambda_1/\lambda_2)$ diverges at the critical point, which is signaled by the degeneracy between the two largest eigenvalues $\lambda_1$ and $\lambda_2$. In a tricritical system, it is known that the third largest eigenvalue $\lambda_3$ also plays an important role at the tricritical point and the first-order transition points where the three largest eigenvalues are degenerate~\cite{Beale1986,Xavier1998,Rikvold1983,Derrida1983,Herrmann1984a,Beale1984,Bartelt1986}. The quantity $\tilde{\xi}_L = 1/\ln(\lambda_1/\lambda_3)$ was proposed primarily to capture this degeneracy~\cite{Rikvold1983,Derrida1983,Herrmann1984a,Beale1984}. This quantity has been often called as the second correlation~\cite{Derrida1983,Herrmann1984a} or persistence length~\cite{Beale1986,Beale1984} which was interpreted as a length scaling of a domain of the disordered phase along the infinite strip whose the scaling behavior is related to the surface tension at the interface~\cite{Rikvold1983}. 

We consider the strip width up to the size of $L=18$ sites which corresponds to the transfer matrix of the size $3^L \times 3^L$ where $3^{18} = 387420489$. In the previous TM calculations, the systems were examined up to $L=10$ with the periodic bound conditions~\cite{Beale1986} and $L=14$ with the helical boundary conditions~\cite{Xavier1998}. While full diagonalization is prohibited for such large matrix, we obtain the cluster of the largest five eigenvalues by using the thick-restart Lanczos method~\cite{Wu2000,Wu1999} which we have implemented with $64$-bit integers to handle a large state index. The restarting scheme can systematically find a few largest eigenvalues by keeping a block of orthogonal basis vectors which is iterated by restarting the Lanczos processes until a desired number of eigenvalues are converged (for details, see Ref.~\cite{Wu1999}). We keep $15$ basis vectors to have five largest eigenvalues converged. Of these, the larger three are used for our analysis while the smaller two are just verified to be well separated from the larger three by a finite gap. Employing such restarting scheme allows much more efficient and reliable numerics than the simple power or Lanczos method when targeting nearly degenerate multiple eigenvalues which are precisely expected in the first-order transitions. 

In the largest case of $L=18$, it takes about $30$ minutes to get one set of eigenvalues when $16$ threads of dual Xeon E5 2.8 GHz are used for the OpenMP parallelization in the sparse matrix-vector multiplication. Comparing with the huge time cost required for the two-parameter Wang-Landau density of states~\cite{Kwak2015}, the transfer matrix is very competitive in terms of overall cost in computational time because the different areas of the phase diagram are searched over in parallel with a computing cluster.

\section{Identifying the first-order transition points}
\label{sec:result1}

\begin{figure}[t]
\includegraphics[width=0.48\textwidth]{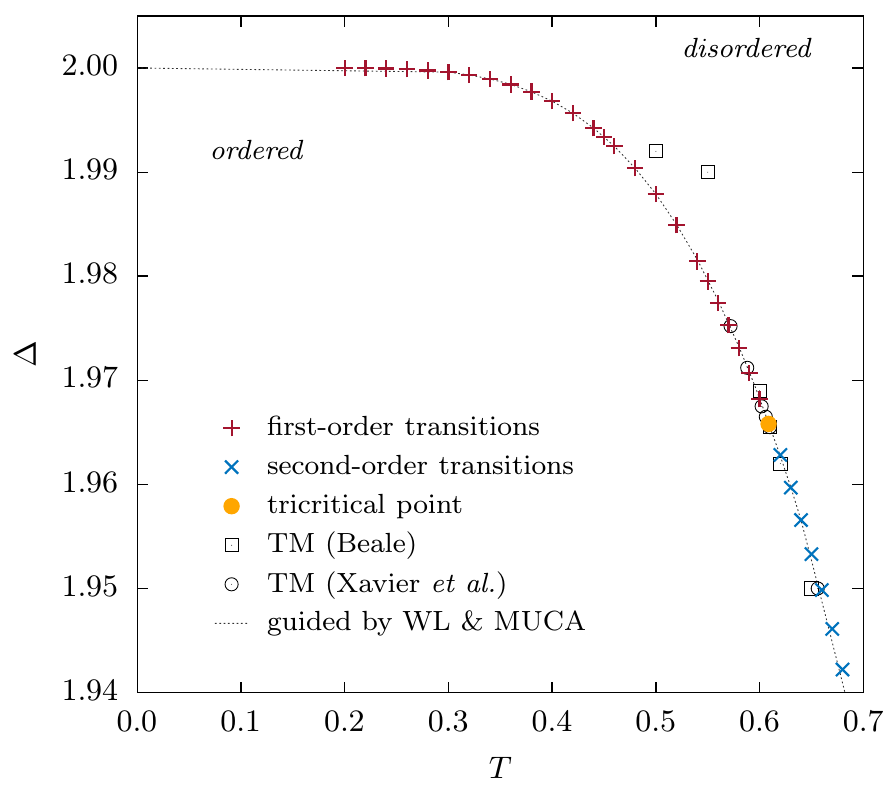}
\caption{Phase diagram of the 2D Blume-Capel model near the tricritical point. The phase coexistence curve in the first-order area of $T \le 0.5$ is determined by the vanishing gap between the largest eigenvalues, and otherwise are the points determined by the finite-size-scaling analysis of correlation length. The error bars are not given because they are smaller than the symbol size (see Table~\ref{tab1}). The previous estimates of the transition points using the transfer matrix method~\cite{Beale1986,Xavier1998} (empty symbols) are given for comparison. The dotted line is guided by the dataset of the recent Wang-Landau~\cite{Kwak2015} and multicanonical~\cite{Zierenberg2017} simulations and the exact point of the boundary at $T=0$ given as $\Delta=z/2=2$ where the coordination number $z=4$.}
\label{fig1}
\end{figure}

In this section, we present the determination of transition points from our TM calculations and show the comparison with the previous TM results~\cite{Beale1986,Xavier1998} and MC estimations of the WL~\cite{Kwak2015} and MUCA~\cite{Zierenberg2017} simulations (see Fig.~\ref{fig1} and Table~\ref{tab1}). The finite-size-scaling analysis of the correlation length is revisited to examine the difference found between the early TM calculation~\cite{Beale1986} and the recent MC estimates of the first-order transition points. At low temperatures, the spectrum of the largest eigenvalues, providing so-called ``free energy levels"~\cite{Privman1983a}, is investigated to characterize the signature of the first-order transitions.  

\subsection{finite-size-scaling analysis of correlation length}

\begin{figure*}[t]
\includegraphics[width=0.95\textwidth]{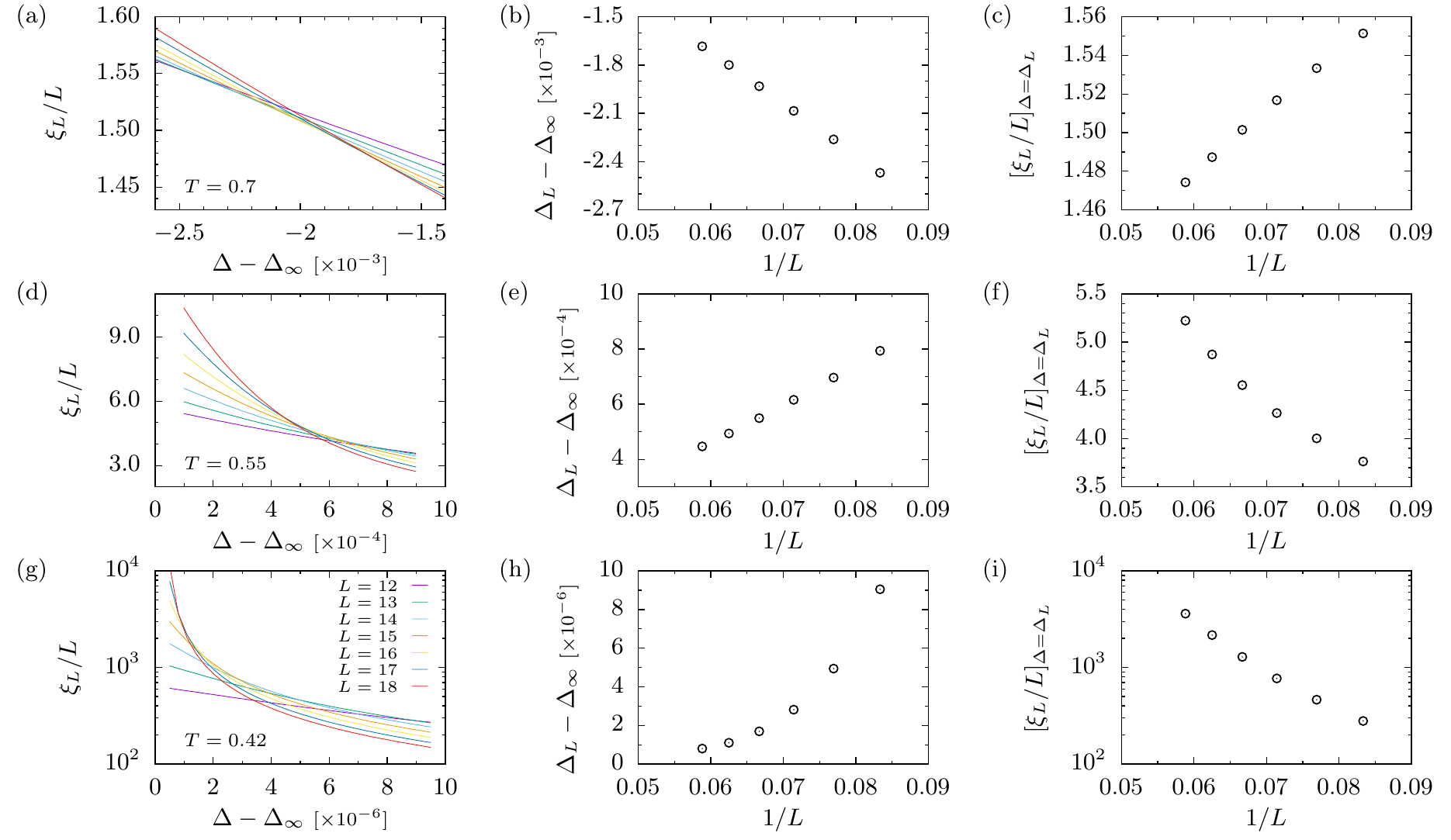}
\caption{Finite-size-scaling analysis of correlation length. Attempts to locate transitions points from the crossing of the scaled correlation length $\xi_L/L$ are demonstrated at the selected temperatures of $T=0.7$ [(a)-(c)] in the second-order area, $T=0.55$ [(d)-(f)] and $T=0.42$ [(g)-(i)] in the first-order area. The crossing point between the curves of $L$ and $L+1$ is denoted by $\Delta_L$, and $\Delta_\infty$ is given by its power-law extrapolation to the limit of infinite $L$ (see Table~\ref{tab1}). }
\label{fig2}
\end{figure*}

Let us begin with a brief review of the finite-size-scaling analysis of the correlation length to determine the critical points. The finite-size-scaling ansatz for the correlation length is typically written in the temperature axis as $\xi_L \approx L \mathcal{Q}(tL^{y_t})$ where the scaling variable $t \equiv (T-T_c)/T_c$ indicates deviation from the critical line, and $\mathcal{Q}$ is a universal function. One can write the similar finite-size scaling ansatz for $\Delta$ as well when approaching the critical line in the $T-\Delta$ plane along the $\Delta$ axis at fixed $T$. Following the work by Beale~\cite{Beale1986} done near the tricritical point, we choose to go along the $\Delta$ axis at given $T$, and then the critical point $\Delta_c (T)$ can be determined by finding a universal crossing point of the curves of $\xi_L / L$ plotted for different $L$'s if the strip width $L$ is large enough.   

However, in practice, for finite systems with small strip widths accessible in the TM calculations, the crossing does not occur exactly at a single common point because of the strong finite-size correction. This has been commonly dealt with by performing the power-law extrapolation to locate the critical point in the thermodynamic limit (for instance, see Refs.~\cite{Derrida1982,Blote1982,Barber1983,Privman1983b,Herrmann1984,Luck1985,Blote1988,Kimel1992,deQuerioz2000}). At a given $T$, the crossing point $\Delta_{L,L^\prime}$ can be determined between the two curves of different strip widths $L$ and $L^\prime$ by solving  
\begin{equation}
\xi_L (\Delta_{L,L^\prime}) / L = \xi_{L^\prime} (\Delta_{L,L^\prime}) / L^\prime.
\label{PR}
\end{equation}
The sequence of $\Delta_{L,L^\prime}$ asymptotically approaches $\Delta_\infty$ which becomes the critical point $\Delta_c$ in the limit of infinite $L$ and $L^\prime$. The convergence has been well studied for several choices of $L$ and $L^\prime$ in the previous works~\cite{Derrida1982,Blote1982,Barber1983,Privman1983b,Herrmann1984,Luck1985}. Following the derivation in Ref.~\cite{Derrida1982} and the review in Ref.~\cite{Luck1985}, the finite-size correction can be written as 
\begin{eqnarray}
\xi_L (\Delta_c) & = & A_0 L ( 1+ A_1 L^{-w} + \cdots ) , \\
\frac{d\xi_L}{d\Delta} \Big\vert_{\Delta_c} & = & B_0 L^{1+y} (1 + B_1 L^{-w}  + \cdots) ,
\end{eqnarray}   
which is used to expand Eq.~(\ref{PR}) around $\Delta_c$ up to the leading order, linearizing an equation for $\Delta_{L,L^\prime}$. For the particular choice of $L^\prime = L+1$, the convergence of $\Delta_L \equiv \Delta_{L,L+1}$ is then shown to be the power-law type as
\begin{equation}
\Delta_{L} - \Delta_c \propto L^{-\tilde\omega} 
\end{equation}
where $\tilde\omega = w+y$~\cite{Derrida1982,Privman1983b,Luck1985}. The extrapolation for $\Delta_\infty$ can be performed for every three data points~\cite{Kimel1992}, which can also be done iteratively~\cite{Blote1982,Blote1988}. Our dataset of $\Delta_\infty$ is the estimate averaged over the possible set of the three-point extrapolations with one point is fixed at $\Delta_{L=17}$, and the uncertainty is specified by their maximal deviation.

This finite-size-scaling analysis of correlation length assumes the criticality of phase transition, and  in principle, it is not applicable to the first-order phase transitions where such ansatz does not hold. However, in practice with small systems, finding the crossing point has been often applied to the area near the tricritical point where the first-order signature is weak~\cite{Beale1986,Rikvold1983}. Here, we examine the temperature range of the applicability of such finite-size-scaling of correlation length, which includes the re-tests of the transition points at $T=0.5$ and $0.55$ that indicated the particular difference between the previous TM calculations~\cite{Beale1986} and the recent MC estimates~\cite{Kwak2015,Zierenberg2017}. 

Figure~\ref{fig2} presents the difference in the finite-size scaling behavior of the scaled correlation length $\xi_L /L$ between the second- and first-order transitions. At $T=0.7$ where the second-order transition occurs, the finite-size correction appears as the power-law convergence of the crossing point $\Delta_L$ with the exponent $\tilde{\omega}$ close to unity. We have observed that $\tilde{\omega}$ increases with $T$ in the examined critical area: for instance, $~\tilde\omega \approx 2.7$ is measured at $T=1.0$. Also, the value of $\xi_L /L$ converges to a finite value as hypothesized in the finite-size scaling ansatz. On the other hand, in the first-order transitions, as shown for the example cases of $T=0.55$ and $0.42$ in Fig.~\ref{fig2}, the scaled correlation length $\xi_L /L$ eventually diverges in the thermodynamic limit, which becomes more evident as $T$ gets lowered to go deeper into the first-order area.  

While the finite-size scaling ansatz is certainly not usable in the first-order transitions, the same extrapolation strategy with crossings between the curves of adjacent $L$'s still provides a reasonable estimate of the transition point in practice unless it is too far from the tricritical point. We obtain $\Delta_\infty = 1.97950(3)$ at $T=0.55$ and $\Delta_\infty = 1.98786(1)$ at $T=0.50$ which correct the previous TM estimate~\cite{Beale1986} of $1.99$ and $1.992$, respectively, and now are precisely on the curve of the WL~\cite{Kwak2015} and MUCA~\cite{Zierenberg2017} data (see Fig.~\ref{fig1}).

This is because of the rapid growth of $\xi_L$ indicating the vanishing gap between the eigenvalues, leading to the crossing points increasingly getting closer to the point of divergence as the strip width increases (see Fig.~\ref{fig2}(g)).
However, at the further lower temperatures, it becomes practically hard to locate the crossing point $\Delta_L$ as already implied in the marginally successful case of $T=0.42$.  Thus, we estimate the first-order transition points directly from the vanishing gap of the eigenvalue spectrum, which extends the accessible area to much lower temperatures as we demonstrate in the followings.

\subsection{free-energy-level spectrum}

\begin{figure}[t]
\includegraphics[width=0.48\textwidth]{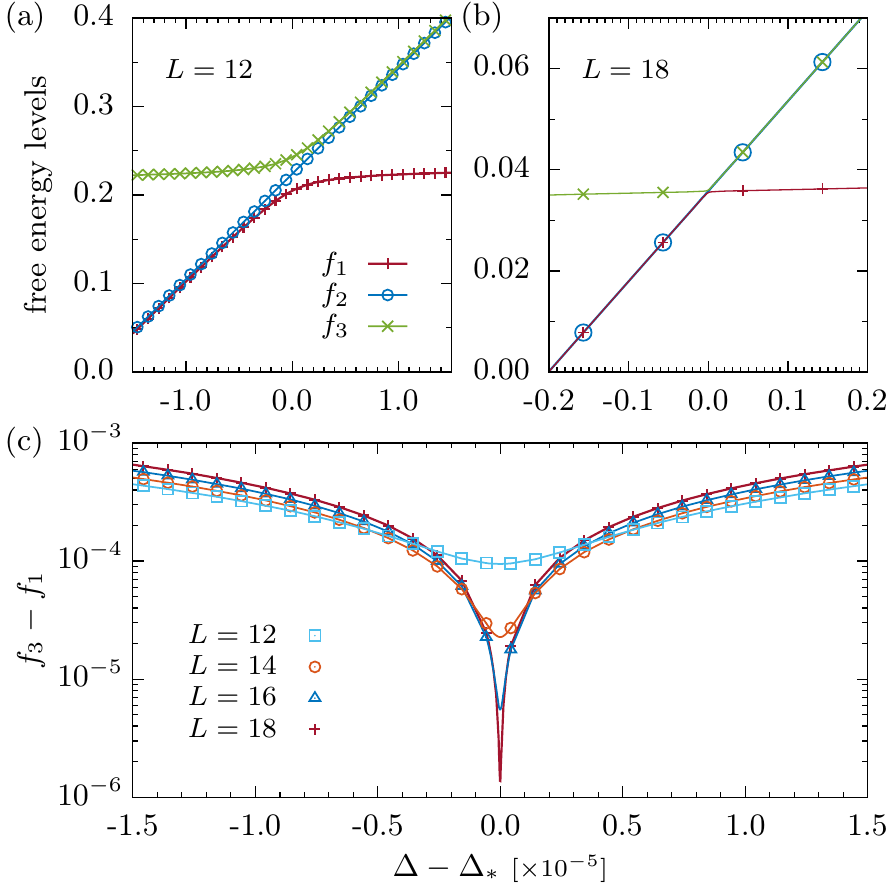}
\caption{Free energy levels and spectral gap across the first-order transition. The lowest three free energy levels $f_i \equiv - \ln \lambda_i$ corresponding to $\lambda_1 > \lambda_2 > \lambda_3$ are plotted at $T=0.4$ for (a) $L=12$ and (b) $L=18$ around the determined transition point $\Delta_*=1.99681357$.  The free energy levels are multiplied by $10^{3}$, and an offset is given for optimal visualization. The data points are computed in the step size of $10^{-8}$  in $\Delta$ near $\Delta_*$. In (c), the gap between $f_1$ and $f_3$ is examined with increasing $L$ to characterize the avoided crossing at $\Delta_*$. }
\label{fig3}
\end{figure}

The free energy level can be defined as $f_i = - \ln \lambda_i$ which is useful to examine how the equilibrium and metastable phases change across the first-order phase transition. In the tricritical Ising-like systems~\cite{Beale1986,Xavier1998,Rikvold1983,Bartelt1986}, the coexistence of two ordered and one disordered phases can be illustrated by the abrupt spectral change of the largest eigenvalues and thereby the free energy levels. While this spectral feature has been demonstrated in the 2D lattice gases~\cite{Rikvold1983,Bartelt1986}, to our knowledge, the anticipated exponential system-size scaling of degeneracy has not been explicitly shown in the ferromagnetic 2D BC model. 

Figure~\ref{fig3} displays the behavior of the lowest three free energy levels at $T=0.4$ calculated as a function of $\Delta$ where the abrupt change the equilibrium free energy of the lowest one is clearly indicated at $\Delta = \Delta^*$. For the disordered phase at $\Delta > \Delta^*$, the single ground level is far from the other two while at $\Delta < \Delta^*$, the lowest levels are doubly degenerate as two ordered phases coexist. The free energy level change at $\Delta = \Delta^*$ gets increasingly sharp as $L$ increases, allowing an accurate determination of the first-order transition point.

The free energy level crossing looks evident at $L=18$, but the gap is always finite in finite-size systems. Such avoided crossing can be characterized by the system-size scaling of the difference between $f_1$ and $f_3$ shown in Fig.~\ref{fig3}(c). It turns out that the gap at $\Delta^*$ exponentially decreases with increasing $L$. This indicates that although the gap does not become strictly zero for the strip widths examined, a kink in the free energy develops exponentially with increasing strip widths, causing an increasingly sharp jump in the thermodynamic quantities that we will show in the next section.  

Finding such $\Delta^*$ giving the minimum gap can determine the first-order transition point. The exponentially vanishing gap is equivalent to the divergence in the correlation and persistence lengths $\xi$ and $\tilde{\xi}$, and therefore $\Delta_\infty$ previously determined from the crossing points of $\xi/L$ should be consistent with the location of $\Delta^*$. The same strategy of searching for the minimum gap was used in Ref.~\cite{Xavier1998} to locate the first-order transition points. Our data of $\Delta^*$ show excellent agreement with the transition points determined in the recent MC studies~\cite{Kwak2015,Valentim2015,Zierenberg2017}.

At low temperatures, it turns out that finding $\Delta^*$ at the minimum gap does not depend much on strip width $L$ as shown in the example case of $T=0.4$ in Fig.~\ref{fig3}(c). For $T \le 0.4$, we find that the value of $\Delta^*$ does not change within the order of $10^{-8}$ for $L$'s examined. This provides a straightforward determination of the phase coexistence curve, which is certainly advantageous over the Wang-Landau approach that requires very careful mixing-field analysis~\cite{Kwak2015}. Above $T=0.4$, weak system-size dependence does appear, where the extrapolation is performed to determine the transition point in the large $L$ limit. The full quantitative comparison between different methods is listed in Table~\ref{tab1}.

\begin{table*}
\caption{Dataset of the transition points. Listed are the estimates of $\Delta_\infty$ from the finite-size-scaling analysis of correlation length and $\Delta_*$ from the free energy level spectrum analysis. The data given without the extrapolation uncertainty are the cases that do not show any system-size dependence within the order of $10^{-8}$ for the larger strip widths with $L > 14$. The datasets from the previous transfer-matrix (TM) calculations~\cite{Beale1986,Xavier1998}, the Wang-Landau (WL) sampling~\cite{Kwak2015}, simulated tempering (ST)~\cite{Valentim2015} and the multicanonical (MUCA) simulations~\cite{Zierenberg2017} are given for direct comparison. The data for Ref.~\cite{Xavier1998} are extracted graphically from the phase diagram.}
\label{tab1}
\begin{ruledtabular}
\begin{tabular}{d{2}d{9}d{9}d{5}d{5}d{8}d{5}d{9}}
\multicolumn{1}{c}{$T$} & \multicolumn{1}{c}{$\Delta_\infty$} & \multicolumn{1}{c}{$\Delta_*$} & \multicolumn{1}{c}{TM~\cite{Beale1986}} & \multicolumn{1}{c}{TM~\cite{Xavier1998}} & \multicolumn{1}{c}{WL~\cite{Kwak2015}} & \multicolumn{1}{c}{ST~\cite{Valentim2015}} & \multicolumn{1}{c}{MUCA~\cite{Zierenberg2017}} \\ 
\hline
0.20 &              & 1.99999080   &        & &           &        &            \\
0.22 &              & 1.99997468   &        & &           &        &            \\
0.24 &              & 1.99994049   &        & &           &        &            \\
0.26 &              & 1.99987615   &        & &           &        &            \\
0.28 &              & 1.99976577   &        & &           &        &            \\
0.30 &              & 1.99958972   &        & & 1.99960(1) &        &            \\
0.32 &              & 1.99932488   &        & & 1.99933(1) &        &            \\
0.34 &              & 1.99894498   &        & & 1.99895(1) &        &            \\
0.35 &              & 1.99870292   &        & &            &        &            \\
0.36 &              & 1.99842103   &        & & 1.99842(1) &        &            \\
0.38 &              & 1.99772164   &        & & 1.99772(1) &        &            \\
0.40 &              & 1.99681357   &        & & 1.99681(1) & 1.9968 & 1.99683(2)  \\
0.42 & 1.9956617(4) & 1.99566194   &        & & 1.99566(1) &        &            \\
0.44 & 1.994232(5)  & 1.9942306(1) &        & & 1.99423(1) &        &             \\
0.45 & 1.993397(1)  & 1.9933985(1) &        & &            &        &             \\
0.46 & 1.992479(1)  & 1.9924828(1) &        & & 1.99248(1) &        &             \\
0.48 & 1.99036(1)   & 1.990380(1)  &        & & 1.99038(1) &        &             \\
0.50 & 1.98786(1)   & 1.98788(1)   & 1.992  & & 1.98789(1) &        & 1.987889(5) \\
0.52 & 1.98490(2)   &              &        & & 1.98496(1) &        &             \\
0.54 & 1.98142(2)   &              &        & & 1.98157(1) &        &             \\
0.55 & 1.97950(3)   & 1.97967(1)   & 1.99   & &            &        &             \\
0.56 & 1.97744(3)   &              &        & & 1.97766(1) &        &             \\
0.57 & 1.97528(4)   &              &        & &            &        &             \\
0.572 &             &              &        & 1.9752       &        &             \\
0.58 & 1.97308(4)   &              &        & & 1.97323(1) &        &             \\
0.588 &             &              &        & 1.9712 &     &        &             \\
0.59 & 1.97072(5)   &              &        & & 1.97080(1) &        &             \\
0.60 & 1.96820(3)   & 1.96817(1)   &        & & 1.96825(1) &        & 1.968174(3) \\
0.602 &             &              &        & 1.9675 &     &        &             \\
0.606 &             &              &        & 1.9665 &     &        &             \\
0.61 & 1.96539(1)   &              & 1.9655 & 1.9655 & 1.96550(1) & &             \\
0.62 & 1.9628(1)    &              & 1.962  & & 1.96270(1) &        &             \\
0.63 & 1.9596(2)    &              &        & & 1.95980(5) &        &             \\
0.64 & 1.9565(1)    &              &        & & 1.9565(1)  &        &             \\
0.65 & 1.9533(1)    &              & 1.95   & & 1.9534(1)  &        & 1.95273(1)  \\
0.656 &                 &              &        & 1.95 &       &        &             \\
0.66 & 1.9498(1)    &              &        & & 1.9501(1)  &        &             \\
0.67 & 1.9461(5)    &              &        & &            &        &             \\
0.68 & 1.9421(5)    &              &        & &            &        &             \\ 
0.69 & 1.9379(5)    &              &        & &            &        &             \\ 
0.70 & 1.9336(4)    &              & 1.92   & &            &        & 1.93296(2)  \\
0.80 & 1.8789(2)    &              & 1.87   & &            &        & 1.87879(3)  \\
0.90 & 1.8029(1)    &              &        & &            &        & 1.80280(6)  \\ 
1.00 & 1.7027(1)    &              &        & &            &        & 1.70258(7)  \\
\end{tabular}
\end{ruledtabular}
\end{table*}

\subsection{interfacial tension and tricritical point}

\begin{figure}
\includegraphics[width=0.48\textwidth]{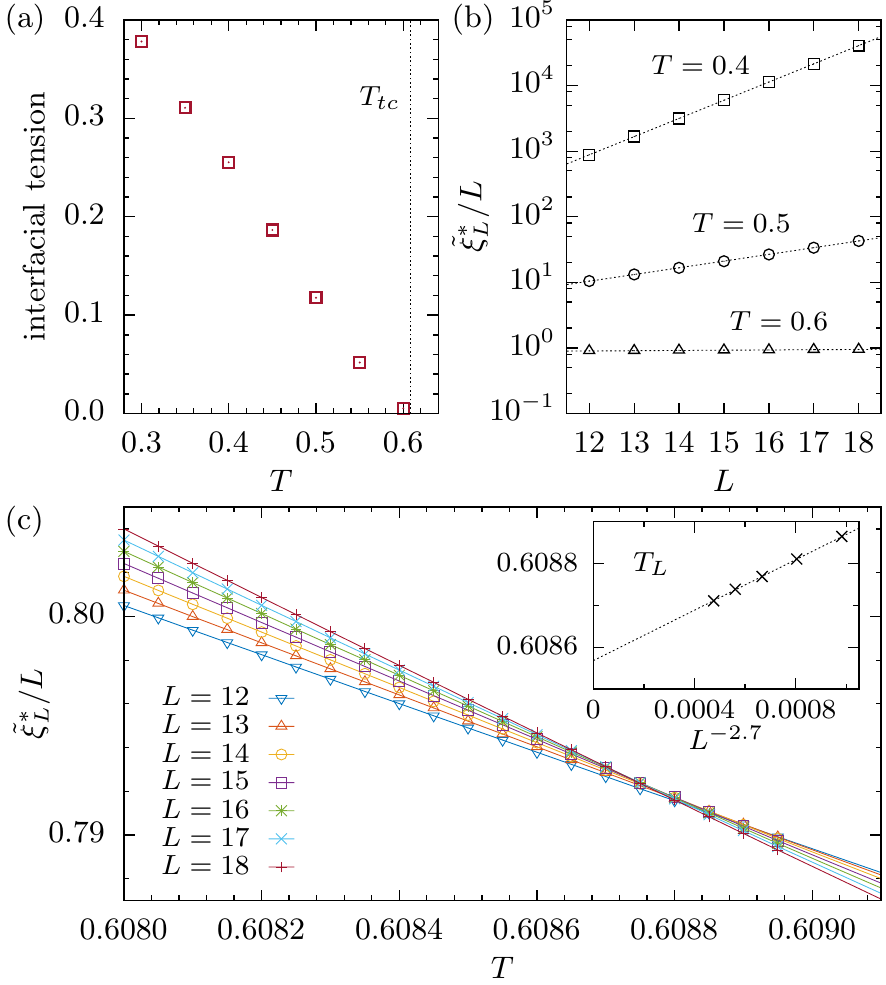}
\caption{Interfacial tension along the phase coexistence curve. (a) Interfacial tension $\sigma$ obtained as a function of temperature. The value is decreased below $0.01$ at $T=0.6$ near the tricritical temperature $T_{tc}$.  (b) Tests for the exponential scaling behavior of the maximum persistence length, $\tilde{\xi}_L^* \sim L \exp(\beta\sigma L)$. (c) Determination of the tricritical point by revising the finite-size-scaling analysis~\cite{Beale1986}. The inset indicates the crossing points $T_L$ extrapolating to $T_{tc} = 0.60858(5)$ leading to $\Delta_{tc} = 1.96582(1)$. }
\label{fig4}
\end{figure}

In the calculations of the free energy levels, we have shown that the gap between $f_1$ and $f_3$ exponentially decreases as the strip width $L$ increases. In terms of the length scale defined previously, this is directly translated into the exponential system-size scaling of the persistence length $\tilde{\xi}^*$. It is known that the asymptotic behavior of the persistence length is related to the interfacial tension between the domains of the phases at the coexistence (see Refs.~\cite{Rikvold1983,Privman1983a} and references therein) of the first-order transition. This is formally written as   
\begin{equation}
\tilde{\xi}^*_L \sim L \exp (\beta \sigma L),
\end{equation}
where $\sigma$ is the interfacial tension.

Along the phase coexistence curve indicated by the peak value of $\tilde{\xi}^*_L$ or the minimum gap between the free energy levels, we present the temperature dependence of the interfacial tension in Fig.~\ref{fig4}(a) which is deuced from the exponential increasing behavior of $\tilde{\xi}^*_L$ as exemplified Fig.~\ref{fig4}(b). It shows that the resulting interfacial tension $\sigma$ diminishes as temperature approaches the tricritical point which is near $T = 0.6$. This is verified by the finite-size-scaling analysis revisited for the tricritical point shown in Fig.~\ref{fig4}(c) although the finite-size correction is still relevant for the strip widths examined. Our estimate of tricritical point $[T_{tc},\Delta_{tc}] = [0.60858(5),1.96582(1)]$ is in very good agreement with the previous estimates of the other numerical studies~\cite{Beale1986,Xavier1998,Landau1986,Wilding1996,Plascak2013,Silva2006,Kwak2015}.

An interesting feature observed in the interfacial tension is that it increases linearly as it goes away from the tricritical point. The larger $\sigma$ implies that the length scaling diverges faster in increasing strip widths. Not only that it explains the sharp determination of $\Delta^*$ with relatively small $L$'s, but it also suggests that a discontinuous signature in a thermodynamic quantity would get more pronounced even in small systems as it goes deeper in the first-order area because of the increased rate of the divergence developing a kink in the free energy. 

\section{Thermodynamic properties: comparison with the Wang-Landau results}
\label{sec:result2}

In this section, we show the consequence of the increasing interfacial tension on the finite-size signature of the first-order transitions in the thermodynamic quantities. For the entropy, non-zero spin density and specific heat, we present how the discontinuity appears along the phase coexistence curve and compare their finite-size influence with the previous Wang-Landau results~\cite{Kwak2015} in the vicinity of the first-order transition points.

In Ref.~\cite{Kwak2015}, the two-parameter WL algorithm was implemented in the standard way to produce the joint density of states (JDOS) for the exact energy level structure of the 2D BC model with full spectral resolution. The WL method is a very efficient tool to access the first-order transitions, but there are some known issues. The error of the WL estimator tends to saturate~\cite{Belardinelli2007,Zhou2008,Liang2006,Schneider2017}, and no rigorous proof exists for convergence in the standard algorithm. The possible issues with the non-integer energy and the multi-range approach have been also pointed out (for instance, see the discussion in Ref.~\cite{Zierenberg2015}), although this was not the case for the one-range implementation with integer energies in Ref.~\cite{Kwak2015}. The most relevant issue here is the tremendous computational time cost of the two-parameter algorithm, severely limiting the accessible size of a system. The one with $48 \times 48$ lattices~\cite{Kwak2015} is currently the largest calculation for the 2D BC model within the two-parameter scheme. The TM calculation only deals with a small width of the strip geometry which, however, is infinite in one direction. Therefore, it may be interesting to compare the appearances of finite-size effects between the present TM calculations and previous WL results~\cite{Kwak2015} on the thermodynamic quantities that are the most readily obtainable from the available WL JDOS.

\subsection{entropy density}

\begin{figure}
\includegraphics[width=0.48\textwidth]{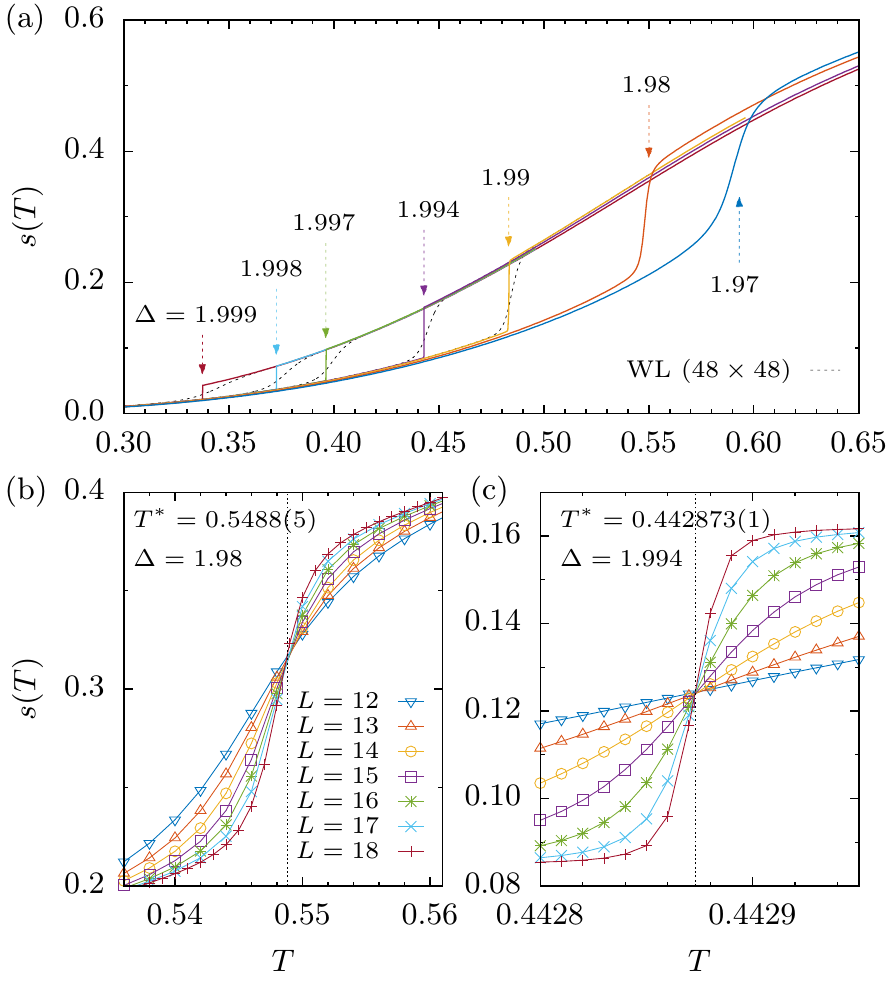}
\caption{Entropy density jump at the first-order transition. (a) The entropy density $s(T)$ is plotted as a function of temperature $T$ at a given crystal field $\Delta$ in the system with the strip width $L=18$. The previous Wang-Landau data in $48 \times 48$ lattices~\cite{Kwak2015} (dotted lines) are given for comparison. The determination of the transition point $T^*$ by the crossing point of $s(T)$ is shown at (b) $\Delta=1.98$ and (c) $\Delta=1.994$. }
\label{fig5}
\end{figure}

Very sharp jumps are identified in the entropy density $s(T)$ shown in Fig.~\ref{fig5}(a) at the transitions in the system with $L=18$. The entropy follows the different curves in the ordered and disorder phases, and it turns out that the two curves become closer to each other as it approaches the zero-temperature point of $\Delta=2$. The entropy difference at the jump accordingly decreases with increasing $\Delta$ as it goes deeper into the first-order area. From the Clausius-Clapeyron relation, this suggest the slope of the phase consistence curve decreases, which is consistent with the obtained phase diagram in Fig.~\ref{fig1}.

While the sharp jump observed provides the evidence of the first-order phase transition, the transition point can be more systematically determined in the test of different strip widths. We find that there exists a well-defined crossing point in the entropy curves of different $L$'s (see Figs.~\ref{fig5}(b) and (c)), and the change around the crossing point gets systematically sharper as $L$ increases, verifying that the transition occurs at the crossing point in the thermodynamic limit. The resulting transition points $T^* = 0.5488(5)$ and $0.442873(1)$ found in the example cases shown at $\Delta=1.98$ and $1.994$, respectively, fall perfectly on the phase coexistence curve in Fig.~\ref{fig1}.

Remarkably at low temperatures, Fig.~\ref{fig5}(a) shows that the TM method outperforms the previous WL calculations~\cite{Kwak2015} in revealing the discrete jumps in the entropy at the first order transitions (see Fig.~\ref{fig5}(a)). In case of the WL data, it seems that the entropy change at the transition occurs more gradually at the higher $\Delta$, which makes the jump harder to be identified. In contrast, in the TM calculations, the jump gets sharper as the transition temperature decreases, demonstrating its accessibility to the deep first-order area of the phase diagram.  

\subsection{non-zero spin population density}

\begin{figure}
\includegraphics[width=0.48\textwidth]{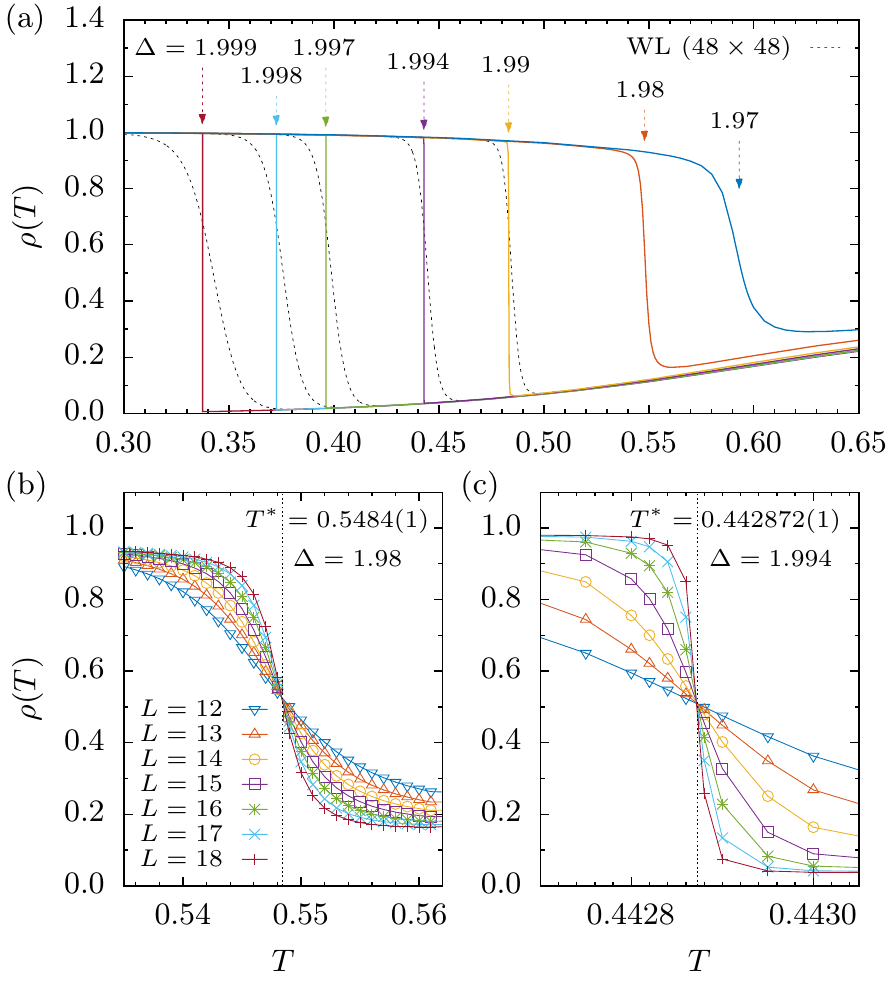}
\caption{Jump of the nonzero spin density at the first-order transition. (a) The population density of nonzero spins $\rho(T)$ is plotted as a function of temperature $T$ at a given $\Delta$. The previous Wang-Landau results~\cite{Kwak2015} (dotted lines) are given for comparison. In (b) and (c), the transition points $T^*$ is determined at the crossing point as done similarly with the entropy density $s(T)$ in Fig.~\ref{fig5}.}
\label{fig6}
\end{figure}

The sharp jumps are evident also in the nonzero spin density $\rho = \langle s^2 \rangle$ as shown in Fig.~\ref{fig6}(a). The change in $\rho$ at the jump monotonically increases with increasing $\Delta$. We observe that the nonzero spin population right after the transition becomes increasingly more depleted as the transition temperature decreases at the higher $\Delta$. While the drop to the finite density of $\rho$ was suggested in the previous WL~\cite{Kwak2015} and the simulated tempering (ST) simulations~\cite{Valentim2015}, our TM calculations clarify the systematic behavior of the drop in $\rho$ in the wide range of the first-order transition line.

The comparison with the recent WL and ST calculations for the non-zero spin density emphasizes again the remarkably small finite-size influence observed in the present TM calculations at the deep first-order area. As seen in the test of the entropy, the nonzero spin density also shows that $\rho$ of the WL calculations loses its sharpness quickly as it goes to the higher $\Delta$ (lower $T$), implying that finite-size effects get stronger. On the contrary, in the TM results, the drop gets sharper as it goes into the first-order area, indicating much suppressed finite-size influence at low temperatures.

The determination of the transition point can be easily done by locating the crossing point in the test of different strip widths as shown in Figs.~\ref{fig6}(b) and (c). It should be noted that the value of $\rho$ at the crossing point is not universal in the TM calculation as opposed to the value of $2/3$ identified in the ST and WL calculations in the isotropic square lattices. This difference is probably due to the setting of the infinite strip geometry imposed in the TM method. Although, in the direct comparison with the curve from the WL data shown in Fig.~\ref{fig6}(a), the crossing points are indeed found between the TM and WL curves at $\rho \simeq 2/3$. This demonstrates that the drops in the TM curves of $L=18$ are already sharp enough very close to the true first-order transition points despite the system being still finite.  

\subsection{divergence of the specific heat}

\begin{figure}
\includegraphics[width=0.48\textwidth]{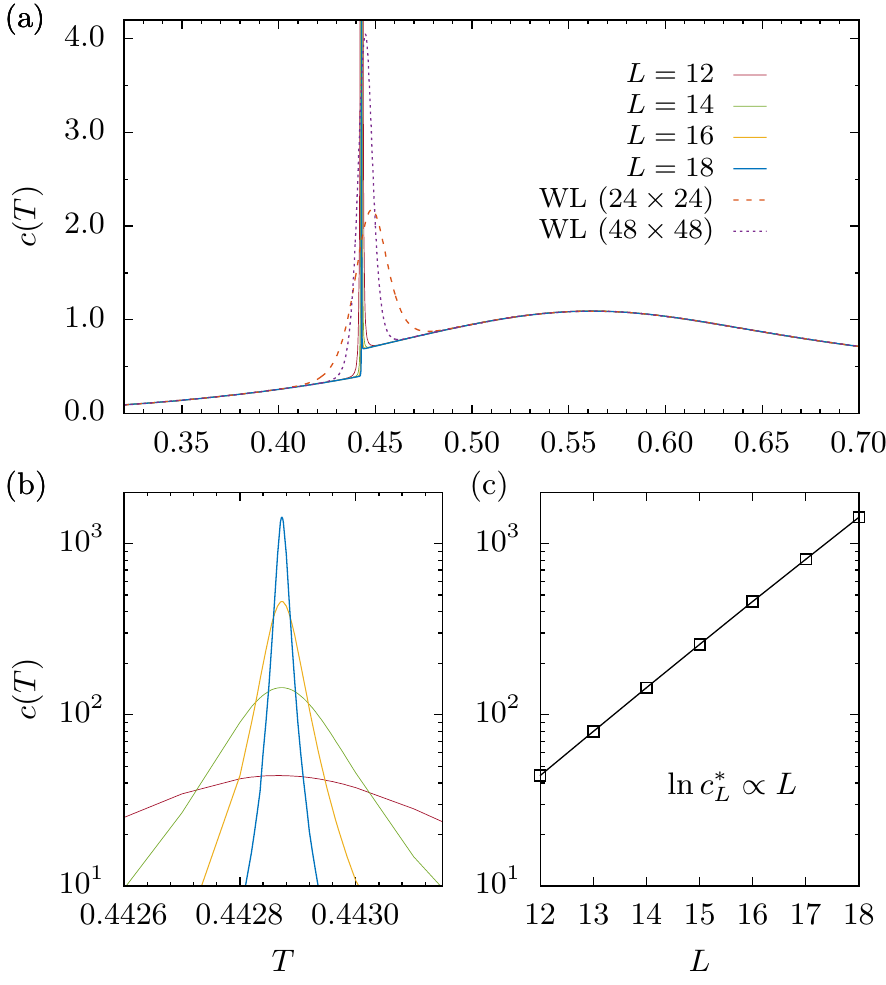}
\caption{Specific heat at the fixed crystal field $\Delta=1.994$. (a) The divergence accompanied with the jump of the specific heat $c(T)$ is found at the transition point. The finite-size behavior is compared with the previous Wang-Landau results~\cite{Kwak2015} for $24\times 24$ and $48\times 48$ lattices. In (b) and (c), the peak value $c^*_L$ of the specific heat is shown to diverge exponentially with increasing strip with $L$.}
\label{fig7}
\end{figure}

Finally, we examine the finite-size influence on the divergence in the specific heat $c(T)$ calculated by the TM method. Figure~\ref{fig7} presents the TM calculations of $c(T)$ at $\Delta=1.994$ showing very clear divergence at the pseudo-transition point $T^* = 0.4428725(5)$ which is in excellent agreement with the estimates from the entropy and nonzero spin density jumps. The peak height grows exponentially with increasing strip width $L$, which is consistent with the exponentially sharpened kink of the free energy indicated in Fig.~\ref{fig3}. In comparison with the previous WL calculations~\cite{Kwak2015}, the feature of the divergence and jump is certainly clearer in the TM calculations, and the Schottky-like hump discussed in the previous WL calculations is well separable without any issue of resolution even in the smallest strip width considered. 

All of the thermodynamic quantities that we have examined consistently demonstrate very clear signature of the discontinuity in the first-order transitions. Although our largest strip width is limited to $L=18$ for the strip width, we find that the influence of the finite-size effects gets more suppressed as it goes deeper into the first-order area, which indicates a clear advantage over the previous two-parameter WL approach where the finite-size influence becomes more pronounced as it goes to the lower temperatures.
            
\section{Conclusions}
\label{sec:conclusions}

We have demonstrated the applicability of the transfer matrix method to the study of the first-order transitions in the 2D Blume-Capel model in square lattices. With strip widths increased up to the size of $18$ sites, we have successfully determined the phase coexistence curve by analyzing the abrupt change in the eigenvalue spectrum of the transfer matrix and also by examining the previous finite-size-scaling analysis of the correlation length. The first-order transition points now obtained with much higher accuracy show excellent agreement with the Wang-Landau~\cite{Kwak2015} and simulated tempering~\cite{Valentim2015} calculations and particularly with the recent highly accurate multicanonical results by Zierenberg, et al.~\cite{Zierenberg2017}. Our transfer matrix calculations have also extended the accessible area to further lower temperatures of the first-order transition. 

In the direct comparison of the thermodynamic properties with the previous Wang-Landau results~\cite{Kwak2015}, we have found that the finite-size influence is much more suppressed in the transfer matrix calculations for the first-order transitions. The transfer matrix calculations of entropy, non-zero spin population, and specific heat show increasingly sharp jumps as the crystal field increases, which contrasts with the Wang-Landau results where the change around the first order transition points becomes rather smoother as it goes deeper into the first-order area. 

The transfer matrix method provides an exact evaluation of the partition function and free energy in the whole range of parameters. Despite the fact that it requires a model-specific formulation for efficient numerics, the accessibility to the first-order transition area certainly benefits from the exact and deterministic nature of the transfer matrix. We argue that the transfer matrix method can be an excellent complementary tool to corroborate the recent development of Monte-Carlo methods on the first-order transitions. We have found that it actually outperforms the previous Wang-Landau results in revealing the discontinuous jumps of thermodynamic quantities in the 2D Blume-Capel model in square lattices.

\begin{acknowledgments}
This work was supported from Basic Science Research Program through the National Research Foundation of Korea funded by the Ministry of Science and ICT (NRF-2014R1A1A1002682) and also from GIST Research Institute (GRI) grant funded by the GIST in 2017.
\end{acknowledgments}

\end{document}